*Research Article*

# Evaluation of TRANSFoRm Mobile eHealth Solution for Remote Patient Monitoring during Clinical Trials

**Jarosław Jankowski,[1,2] Stanisław Saganowski,[1] and Piotr Bródka[1]**

[1]*Department of Computational Intelligence, Wroclaw University of Science and Technology, 50-370 Wrocław, Poland*
[2]*Faculty of Computer Science and Information Technology, West Pomeranian University of Technology, 71-210 Szczecin, Poland*

Correspondence should be addressed to Stanisław Saganowski; stanislaw.saganowski@pwr.edu.pl

Today, in the digital age, the mobile devices are more and more used to aid people in the struggle to improve or maintain their health. In this paper, the mobile eHealth solution for remote patient monitoring during clinical trials is presented, together with the outcomes of quantitative and qualitative performance evaluation. The evaluation is a third step to improve the quality of the application after earlier Good Clinical Practice certification and validation with the participation of 10 patients and 3 general practitioners. This time the focus was on the usability which was evaluated by the seventeen participants divided into three age groups (18–28, 29–50, and 50+). The results, from recorded sessions and the eye tracking, show that there is no difference in performance between the first group and the second group, while for the third group the performance was worse; however, it was still good enough to complete task within reasonable time.

## 1. Introduction

The standard method of collecting PROMs (Patient Reported Outcome Measurement) relies on paper forms that are presented to the patient. A more recent approach uses web or mobile software to assess patient health status and quality of life [1–3]. Electronic monitoring of PROMs allows the health of patients with chronic disease such as diabetes mellitus and gastroesophageal reflux disease (GORD) to be monitored closely, without the need to visit a health institution for each report. In addition, those data can be preprocessed automatically by algorithms which are looking for alarm symptoms and signs and if necessary notify the GP (general practitioner) that the patient needs attention. These features can thus improve the quality of care and the quality of life for patients requiring close monitoring, like elderly people or people suffering from chronic diseases.

Despite the potential benefit of this approach, there are currently no widely accepted standards for developing or implementing PROMs. From time to time, targeted solutions are developed to run a study focused on a specific trial [4].

The presented research shows the evaluation of the new solution in the area of remote patient monitoring during clinical trials via mobile devices, based on the CDISC ODM standard (http://www.cdisc.org/odm). Importance of such solutions grows together with new regulations addressed to medical storage data and new forms of communication with patients. Additionally, such solutions should decrease the cost of randomized controlled trials (RCT) and, what is more important, with fewer visits in the health institution, increase the comfort of the patient. Finally, to the best of our knowledge, TRANSFoRm Clinical Trial Management System is the first working system which enables running any RCT designed with the use of ODM/SDM standard.

Mobile applications designed and implemented within TRANSFoRm project were GCP (Good Clinical Practice) certified and validated with the participation of 10 patients and 3 general practitioners [5]; however, neither of those procedures revealed any substantial evidence on how to improve the mobile applications. That is the main motivation for the additional, quantitative, and qualitative performance evaluation of mobile applications. In this paper, the entire evaluation and its outcomes are presented. The paper is organized as follows: in the next section the related work is presented; Section 3 gives as a brief introduction to TRANSFoRm Clinical Trial Management System for which mobile applications are one of the key components; Section 4 contains general description of TRANSFoRm mobile applications and their functionality; Section 5 includes the complete evaluation of TRANSFoRm mobile applications with the brief description of previous actions (GCP certification and real world validation); finally, the whole paper is concluded in Section 6.

## 2. Related Work

The increasing role of the mobile technologies in various areas related to the healthcare including communication with healthcare institutions [6], access to the health-related information, education and promotion of the healthy lifestyle [7], chronic disease prevention [8], monitoring [9], or medical decision-making [10] is observed. The mobile technologies are transforming healthcare towards the more open systems with better availability [11].

The market of mobile applications targeted to eHealth grows in several directions with solutions provided by independent vendors and institutions. Even though attempts towards the standardized platforms are taken, the standards market is very fragmented. The number of the health-related apps reached in the past years more than 100,000 applications with the main purpose to record, track, and analyse the behaviours or the health data over time [8]. Massively available online applications are targeted to the healthy life styles [12], fitness and physical activity monitoring [13], weight loss programs [14], healthy foods [15], and various other areas.

The application of mHealth for chronic or long-term illnesses care is one of the most significant directions in the health system development over the past years [16, 17]. The recent research includes applications for self-management for diabetes [18] with effectiveness evaluation [19] and the focus on the mobile phones [20]. Dedicated applications are targeted to the asthma patients [21], cancer supportive care [22], or HIV/AIDS care [23].

Apart from massively available applications, the specialized applications are being developed for access to radiology systems [24], supporting the orthopedic decisions [25], anesthetic decisions and processes [26], or monitoring and tracking infectious diseases [27].

Earlier studies mention the need for the better usability and patients confidence during interactions with mHealth systems [28]. Developers of eHealth tools apply the user centred design and usability studies to detect the differences between the end-user needs and the developers' perceptions of the clinical applications [29]. The advantages of involving the users into the process of designing medical applications and technologies is emphasized; however, the barriers are identified related to increased costs and additional time required for the development [30]. A special care and rigorous approach to the development of patient targeted applications should be implemented into the design process [31]. Another issue is the transparency of the medical applications supporting decision-making [10]. Yet another issue is improving the usability for users of different ages. The mobile applications related to the healthcare might need to deliver special functions for the patients of age 50 or more, what was shown in the application for patients with diabetes [32]. The special functions include the screen readers, the ability to adapt the size of control elements, or adjusting the contrast. The study shows moderate to good results for the applications with a small range of functions while the usability of multifunctional apps was evaluated as worse.

While most of the systems offer dedicated solutions for specific area of applications, the problem is the standardization, usability, security, and transparency of data processing algorithms. Individual development of applications for various areas requires additional costs for implementation and applications do not always offer required functionality and convenient access for patients from different target groups and of different ages. The TRANSFoRm project delivers standardized mobile eHealth solutions for remote patient monitoring, mainly during RCTs, but with the possibility of constant patient monitoring. The main goal of the presented study is the evaluation of the proposed solutions in terms of standards and performance with the focus on quantitative approach and eye tracking based analysis. It is the next stage of research presented earlier [5, 33].

## 3. TRANSFoRm Clinical Trial Management System

TRANSFoRm is an EU funded large scale project within the 7th Framework Programme, with three main objectives, which are (1) to facilitate multiple site genotype-phenotype studies, (2) to prototype a diagnostic decision support system linked to Electronic Health Record (EHR) systems, and (3) *to enable multisite, multilanguage, practice-based randomized controlled trials (RCTs)* by integration with existing EHR systems. A core output of the project is the specification and demonstration of a "functional" eCRF (electronic Case Report Form), designed to enable the collection of semantically controlled and standardized data within existing EHR system.

The third objective is based on the clinical research question "does continuous PPI (Proton Pump Inhibitor) differ from on-demand PPI use regarding symptom severity and quality of life [34]"? To answer that question, a multicentre international RCT including 600 GORD patients randomized to continuous or on-demand PPI treatment has been designed [5], EudraCT-number 2014-001314-25.

The functionalities of the TRANSFoRm applications include (1) identifying prevalent and incident cases of GORD, (2) randomizing patients to on-demand or continuous consumption of PPIs, and (3) following these patients using patient mobile or web applications and eCRFs completed by medically qualified personnel at practice visits. The data submitted by the patients using the mobile or web applications are PROMs while the data entered by the clinician using eCRFs are CROMs (Clinician Reported Outcomes Measurement). The task was to build the system which can easily integrate with existing systems, that is, different EHRs, and allow fully conducting multicentre, multilanguage international randomized controlled trial and at the same time make it as easy as possible for the patients and GPs.

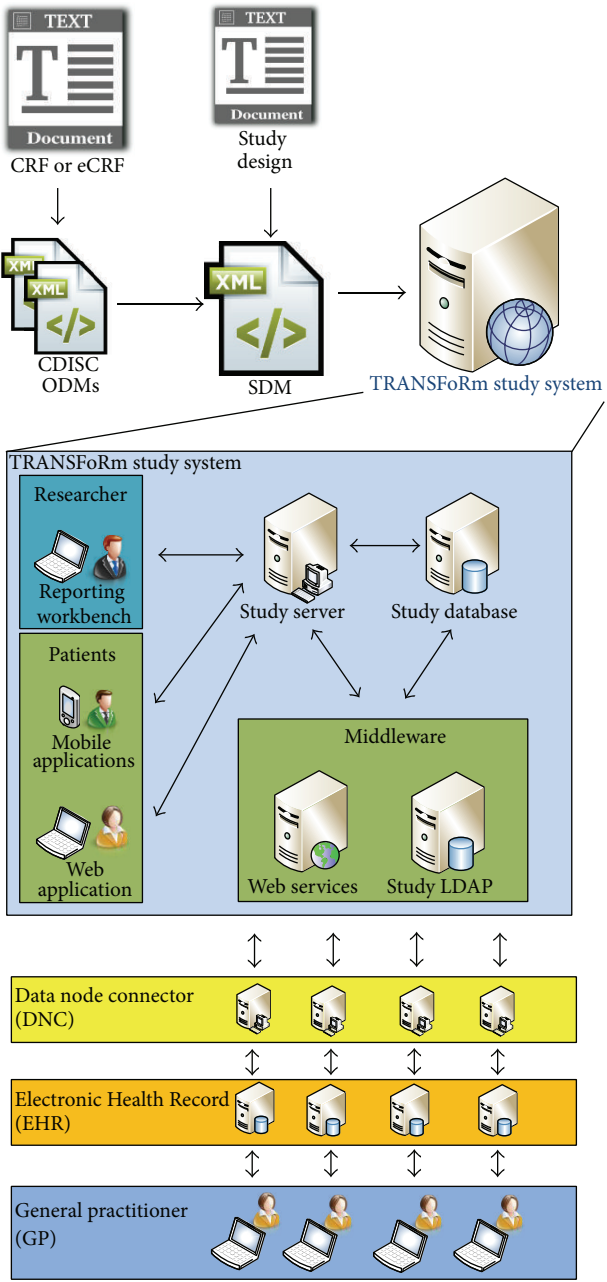

Figure 1: The architecture of the TRANSFoRm Study System.

The TRANSFoRm Study System (TSS) is an electronic platform to collect PROMs and transfer data to the EHR systems. The TSS consists of five major parts (Figure 1):

(i) Study Server (SS), which manages the connection between mobile and web applications and the external parts outside of the TSS.

(ii) Study Database (SDB), which stores all of the information about studies, patients, randomization, and so forth and is used also by the middleware and the Data Node Connector (DNC).

(iii) Web application, which is an application placed on the web server that enables filling out PROMs by the patients and CROMs by the GPs.

(iv) Mobile applications, which are native applications for Android and iOS systems that enable filling out PROMs by the patients.

(v) Middleware, an Enterprise Service Bus, which serves as a connection, authorization, and security layer between TRANSFoRm Study System and the rest of the TRANSFoRm infrastructure.

This paper focuses on the mobile applications and their validation but if the reader is interested in the retest of the system please see [33] and TRANSFoRm project Deliverables.

## 4. Mobile Applications

TRANSFoRm Study System mobile and web applications are designed to enable the study participants to fill out the PROM questionnaires. The mobile applications are available on Android and iOS platforms. The applications are capable of generating human readable version of any questionnaire provided in the ODM (Operational Data Model) standard. The interface was designed with respect to the Android and iOS platforms' guidelines and best practices. Moreover, the applications were built upon the standard system user interface elements; therefore, using the application should be comfortable and intuitive to the patients.

*4.1. Requirements and Availability.* The Android application is compatible with the system version 4.0 and higher while the iOS application requires system version 7.0 or higher. The mobile devices running other platforms are not able to operate the TRANSFoRm Study System mobile applications but can still use the web application through their system web browser. The mobile applications are available on platform specific app markets, the Android application at [35] and the iOS application at [36]. The web application is available at [37].

A user account in the TSS is created for patients enrolled in the clinical research study. A unique user name and password for the mobile and web applications are provided to the patient by an email.

Every patient authorized by the TRANSFoRm Study System can perform the following actions within the applications:

(i) Log in and log out.

(ii) See the list of pending and completed questionnaires.

(iii) Fill out pending questionnaires and send them to the TRANSFoRm Study System.

(iv) Close a questionnaire during the process of filling it out (the current progress will be lost).

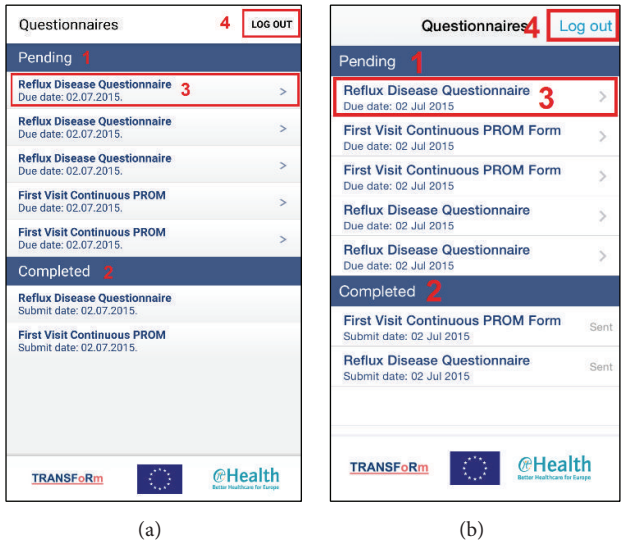

Figure 2: The questionnaires list screen in the Android (a) and iOS (b) applications.

In case of any warnings or errors, the mobile application notifies the patient through the system alert boxes.

*4.2. Questionnaires List Screen.* The "Questionnaires" screen (Figure 2) contains the full list of questionnaires assigned to the patient. The list is divided into pending (area 1) and completed questionnaires (2). The pending questionnaires can be filled out while the completed questionnaires are nonselectable and thus cannot be changed. In order to fill out the questionnaire, the patient has to click on the desired questionnaire (3). The questionnaire assigned to the patient is defined by the clinical researchers and is allocated to the patient in the TRANSFoRm Study System.

The "Questionnaire" screen allows the patient to fill out the pending questionnaire. The example of filling out the Reflux Disease Questionnaire is presented in Figure 3. The patient should perform the following actions there:

(1) Answer all questions on the first screen by selecting the appropriate answer from the list.
(2) Click the NEXT button.
(3) Answer all questions on the second screen.
(4) Click the Send button.

If some answers are missing, the alert box pops out and the questions without answers are marked red as in Figure 4.

The questionnaire can be closed at any time by the Close button. In such case, the current progress will be lost and the empty questionnaire still will be available as a pending questionnaire.

*4.3. Data Safety.* A communication between the components is held over a secure SSL connection and the requests are structured in XML format. Communication with the TRANSFoRm Study System requires a valid session key which is generated every time the patient logs in into the system. For safety reasons, the key is valid for 30 minutes. After that time, the first attempt to communicate with the TRANSFoRm Study System will automatically log the patient out from the mobile or web applications and if the patient wishes to continue working in the TSS system it is required that they log in again.

To ensure the highest possible security and data privacy no data are stored on the patients device, which is the reason why answers cannot be saved and the questionnaire has to be completed at one time point. The completed questionnaires are immediately sent to the TRANSFoRm Study System and stored in the TRANSFoRm Study Database.

*4.4. Generalisability.* The mobile applications were designed to be as flexible as possible in terms of supporting different types of questionnaires and multiple language versions at the same time. The applications are able to generate a human readable version of any questionnaire as long as it is compatible with the ODM standard (http://www.cdisc.org/odm). The applications are equipped with built in logic to parse and display ODM-structured files. Therefore, providing the new type of questionnaire to the patients requires only creating the proper XML document.

Furthermore, the mobile and web applications are able to properly switch the language of the questionnaire depending on the language selected on the patient's device or in the patient's browser. The only requirement is that the appropriate translation is available in the ODM file describing the questionnaire. Currently, the TRANSFoRm study includes four languages: English, Polish, Dutch, and Greek. Adding another language is a simple process; it requires translating the questionnaire into a new language and embedding the translation into the ODM file. An example of translating the "Self-Rated Health" question into Dutch and Polish is presented in Figure 5.

*4.5. ODM User Interface Extension.* The study designer can design the eCRF appearance on different ways, according to his needs. On the other hand, the user interface elements available in the Android and iOS platforms are very specific when it comes to how they look and react. In order to satisfy the study designer needs and ensure the intuitive usage of the applications, the novel ODM UI extension has been proposed.

When an eCRF is created, the additional attribute *QuestionType* can be added to *ItemDef* object (which represents a single question within the questionnaire). The *QuestionType* attribute will indicate how the answers to this particular question will be displayed to the users; for example, adding *QuestionType="DropDown"* will cause displaying the answers to that question as a drop down list on both mobile platforms. The study designer can choose one of the six predefined types for the *QuestionType* attribute (see Table 1). It is the study designers' responsibility to use the *QuestionType* attribute properly. In the case of inappropriate use, the eCRF might be rendered incorrectly. If the attribute *QuestionType* will not be

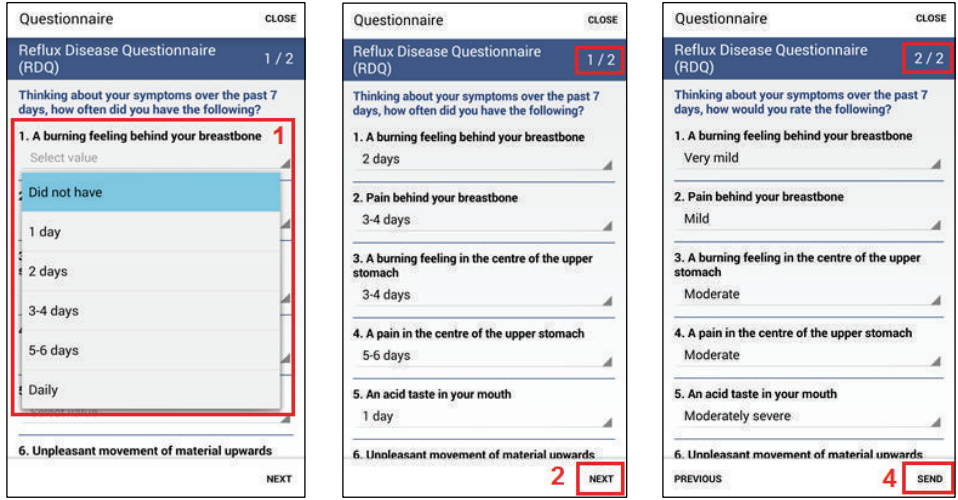

Figure 3: The example of filling out the Reflux Disease Questionnaire in the Android application.

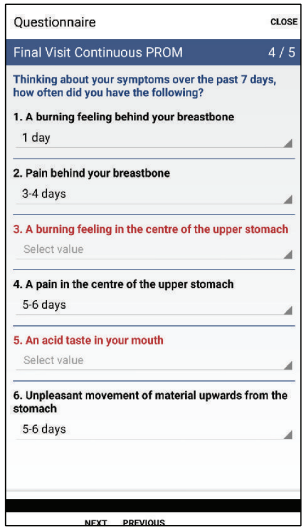

Figure 4: The example of missing answers.

provided, the default question type will be chosen using the following rules:

(i) If *ItemDef* contains attribute *DataType*="*date*", then *QuestionType*="*DatePicker*".

(ii) Else if *ItemDef* contains a *CodeListRef* tag, then *QuestionType*="*DropDown*".

(iii) Else *QuestionType*="*InputField*".

When deciding which user interface elements will be used, the idea was to keep standards persistent on both platforms, this way the users will be familiar with the interface and will know how to interact with it. Therefore, the same question may look different on each platform. What is more, the same question may even look different on two Android devices (depending on the system version). The goal is to utilize user experience with his own mobile device.

## 5. Evaluation

Entire TRANSFoRm Study System including mobile applications has been validated in three steps. First *(GCP certification)*, the applications have been GCP (Good Clinical Practice) certified [5]. Second *(Validation)*, the application has been tested and validated by test patients and GPs [5]. Third *(Evaluation)*, the applications are being evaluated in a randomized controlled trial comparing the full TRANSFoRm system with manual patient recruitment, a web based and paper based PROM collection. 600 patients from four different countries are participating in the study. The full outcome of RCT evaluation will be known in June 2016.

Unfortunately, neither GCP certification nor validation revealed any substantial evidence on how to improve the mobile applications. That is why one more midstep, that is, quantitative and qualitative performance evaluation, was added for mobile applications between the second (validation) and the third steps (evaluation).

*5.1. GCP Certification and Validation.* Government rules, regulations, and guidance documents contain specific requirements for computerized systems. One of the most important certifications for medical software is the GCP. It was vital for the TSS to be GCP certified. To satisfy this and validate the TSS, two studies were conducted to satisfy the GCP requirements. First, the TRANSFoRm software was installed in three selected practices in Poland and 10 patients at these practices were recruited to test the application to fill out the PROMs. Second, data for 10 simulated test patients were inserted into the mobile and web application to test the system.

Overall, the patient's experiences using the applications were positive, with 6 out of 10 patients preferring the application over completing the questionnaire on paper. The only concern from the patients was the length of the questionnaires and that it took them up to 14 minutes to complete it. However, since the number of questions and their

```xml
<ItemDef DataType="text" Length="40" Name="PAST_SYMPTOM01_FREQ Item" OID="ID.PAST_SYMPTOM01_FREQ">
  <Question>
    <TranslatedText xml:lang="en">A burning feeling behind your breastbone</TranslatedText>
    <TranslatedText xml:lang="nl">Een brandend gevoel achter het borstbeen</TranslatedText>
    <TranslatedText xml:lang="el">Αίσθημα καούρας πίσω από το στέρνο</TranslatedText>
    <TranslatedText xml:lang="pl">Uczucie pieczenia za mostkiem</TranslatedText>
  </Question>
  <CodeListRef CodeListOID="CL.SYMPTOM_FREQ"/>
</ItemDef>
```

Figure 5: The example of translating a single question in the ODM format.

Table 1: The predefined types of the *QuestionType* attribute.

| QuestionType | UI control name | Android component | iOS component |
| --- | --- | --- | --- |
| InputField | Input field | EditText | UITextField |
| DatePicker | Date picker | DatePicker | UIDatePicker |
| RadioButton | Radio button | RadioButton | UITableView |
| YesNo | Switch control | Switch | UISegmentedControl |
| DropDown | Drop down list | Spinner | UITableView (new screen) |

division into sections depends on the ODM files, no changes to mobile and web application interfaces have been made. For more detailed results, please see [5].

*5.2. Research Methodology.* The main goal of this analysis was to verify the performance of the designed application with the focus on quantitative metrics related to the usage of questionnaires. The first system usage was the TRANSFoRm RCT for GORD gastroesophageal reflux disease; however, the entire system is designed in such a way that it can run any randomized control trial designed using CDISC ODM/SDM standards.

For the purpose of this research, one exemplary questionnaire from the original GORD study was used. It contains 15 questions divided into five sections with a varying number of questions: 3, 2, 1, 6, and 3, respectively, so inter- and intrasection performance can be verified. The structure of the questionnaire and questions is presented in Table 2. For questions Q1 and Q5, a virtual keyboard was used for numeric answer entry. For all the other questions, lists with selectable answers were displayed.

A total number of 17 users (13 males and 4 females) from three different age ranges were invited to participate in the study: five users in group A with age in the range of 18 to 28 (mean = 24.8, min = 21, max = 28, and median = 25), nine users from group B with age in the range of 29 to 50 (mean = 36.38, min = 31, max = 44, and median = 35.5), and three users (group C) with age above 50 (mean = 63.67, min = 57, max = 69, and median = 65). The sample size is fitting for usability and it is consistent with earlier research showing that even a small sample of 5 participants can detect 80%–85% of the usability problems [38, 39]. The study uses both qualitative and quantitative approaches and incorporates factors directly contributing to ease of learning and effectiveness. The application was used with the typical scenarios based on answering questions within electronic questionnaire following the general approach proposed by Barnum [40]. Research follows ISO-9241 product efficiency defined as "*resources spent by user in order to ensure accurate and complete achievement of the goals*". For software products and information systems, time spent by the user, in order to achieve the goals, is the key measure. Overall, efficiency can be calculated as the user effectiveness divided by the time spent by the user within sections of application [41].

The application was displayed on the specially mounted 15.6 inch touch screen with the working size of the application scaled to 7.2 inch (4 × 6 inch). It was equivalent to the biggest smartphones (Samsung Galaxy Mega 2 7.0, ASUS PhonePad 7, and Huawei P8 Max) and was only 2 inches bigger than the most popular smartphones, like iPhone 6S (5.5) and Samsung Galaxy S6 edge+ (5.7) and S7 edge (5.5). The user sessions were fully recorded in the form of video stream and further analysed with focus on the navigation between parts of the application, mouse movements, and the selection of the questions and the answers within the application. The observations and notations about the individual task performance were applied during the analysis of the video material.

The eye tracking allowed for the monitoring of visual activity and for checking on how the content of the questionnaire is processed. The eye tracking measurement was conducted with a 60 Hz sampling rate using Gazepoint GP3 eye tracker. The functioning of the device was explained to every participant before the experiment. Every participant's position was set to minimize individual differences in head placement. After setting a proper angle and distance, the calibration process took place. During the calibration procedure, which was 15 seconds long, the participants' task was to observe 9 points arranged on the screen. After the proper calibration, the participants were acquainted with the procedures during the experiment. The analysis of the recorded sessions from the eye tracker was based on the analysis of fixations and saccades. The fixation denotes a phenomenon of aiming the inner part of retinas of both eyes at the object being watched. This action lasts from 0.15 to 1.5 seconds. The measurement of the fixation may refer to the area of interest (AOI) [42], timespan [43], fixation repeatability [44], the percentage share of fixation in AOI [45], or spatial density [46]. The movement of the eye between two fixations is represented by saccades and they occur 4–6 times per second and last approximately from 0.03 to 0.06 seconds. Similar to fixation, a saccade can be interpreted differently depending on the context [47]. During the research, several factors were measured including the user's time that was spent on each part of the questionnaire, the number of views to the defined area of interest (AOI), focus time on the analysed sections, time to the first view of the AOI after the section of questionnaire was fully loaded, and the number of repeated visits to the specific AOI.

Table 2: The content of questions used in the study.

| Section | ID | Question content |
|---|---|---|
| S1 | Header | PROM Demographics |
|  | Q1 | How many persons (excluding yourself) currently live in your household? Number of persons: |
|  | Q2 | How would you describe your current occupation or employment status? |
|  | Q3 | What is the highest level of education that you achieved? |
| S2 | Header | Smoking |
|  | Q4 | Do you currently smoke? |
|  | Q5 | If yes, number of cigarettes/day? |
| S3 | Header | Self-Rated Health |
|  | Q6 | How would you rate your general health status |
| S4 | Header | Thinking about your symptoms over the past 7 days, how often did you have the following? |
|  | Q7 | A burning feeling behind your breastbone |
|  | Q8 | Pain behind your breastbone |
|  | Q9 | A burning feeling in the centre of the upper stomach |
|  | Q10 | A pain in the centre of the upper stomach |
|  | Q11 | An acid taste in your mouth |
|  | Q12 | Unpleasant movement of material upwards from the stomach |
| S5 | Header | Are you suffering from? |
|  | Q13 | Unintentional weight loss |
|  | Q14 | Difficulties swallowing |
|  | Q15 | Anemia |

The analysis of the results is divided into two parts. First, the video stream was analysed. The annotations were added to each subtask and the time of finishing each stage of the questionnaire was retrieved. The main goal was to evaluate the performance of using the application and identify individual drawbacks. The detailed analysis of behaviours within each part of questionnaire for each group of participants was performed. Secondly, eye movement patterns were analysed from the data recorded with the eye tracker and the behaviours were analysed in each age group.

*5.3. Quantitative Analysis of Task Performance.* For users in each group, the total time to complete the task was obtained from the recorded session. For group A, the mean time to finish the task was 149.81 seconds. The users in group B finished the task with an average time of 145.43 seconds, and in group C the mean time was 227.07 seconds. The result for group C was 1.52 times higher than for the users in group A and 1.56 times higher than for the users in group B. For the statistical verification of the results, a Mann-Whitney $U$ Test with continuity correction was used for comparing times achieved for the users in each group. No statistical difference was observed between the users from groups A and B, while the differences between the users from groups C and A as well as C and B were significant at $p < 0.05$ (see Table 3).

The process of the application usage was divided into 22 stages: starting with the *Log in* to the system at Stage 1 and finishing with the *Log out* from the system at Stage 22. For each stage, the time was retrieved from the recorded video session with the annotations and mark-ups added. The *Log in* time was measured from the time when the login screen, with the required login name and password, was presented to the user after the explanation of the experiment. After the *Log in* stage, the user had to select the proper questionnaire. The time when the first part of the application with three questions was loaded is represented by the time in Stage 2. The time for answering the first question on each screen (Q1, Q4, Q7, Q8, and Q14) was measured from the time when section was fully loaded until the answer was selected for the question. Times for all the other questions were based on the intervals between finalizing the answer on a previous question to answering the current question. For example, the time for Q2 was based on the difference between the time when the answer for Q2 was selected and the time when Q1 was answered. The average times for finishing each question for all user groups A, B, and C are presented in Table 4.

Table 3: Intragroup comparison with Mann-Whitney $U$ Test.

| G1 versus G2 | Rank G1 | Rank G2 | $U$ | $Z$ | $p$ value |
|---|---|---|---|---|---|
| B versus A | 61 | 44 | 16 | −0.80 | 0.42 |
| C versus A | **21** | **15** | **0** | **2.09** | **0.04** |
| C versus B | **31** | **47** | **2** | **2.03** | **0.04** |
| G1 versus G2 | $Z$ adjusted | $p$ value | Valid N G1 | Valid N G2 | Rank G2 |
| B versus A | −0.80 | 0.42 | 9 | 5 | 0.44 |
| C versus A | **2.09** | **0.037** | **3** | **5** | **0.04** |
| C versus B | **2.04** | **0.042** | **3** | **9** | **0.04** |

Table 4: Average times in seconds for participants from groups A, B, and C for each question.

| Stage | ID | Average time | | | Mean | Intragroup relation | | |
|---|---|---|---|---|---|---|---|---|
| | | Group A | Group B | Group C | For all groups | B versus A | C versus A | C versus B |
| 1 | Log in | 21.01 | 24.37 | 25.47 | 23.58 | 1.16 | 1.21 | 1.04 |
| 2 | S1 | 11.60 | 10.41 | 16.04 | 11.76 | 0.90 | 1.38 | 1.54 |
| 3 | Q1 | 8.84 | 11.74 | 17.41 | 11.89 | 1.33 | 1.97 | 1.48 |
| 4 | Q2 | 11.26 | 7.26 | 16.75 | 10.11 | 0.64 | 1.49 | 2.31 |
| 5 | Q3 | 10.95 | 13.51 | 20.91 | 14.06 | 1.23 | 1.91 | 1.55 |
| 6 | S2 | 11.98 | 5.59 | 8.71 | 8.02 | 0.47 | 0.73 | 1.56 |
| 7 | Q4 | 5.17 | 4.11 | 6.46 | 4.84 | 0.79 | 1.25 | 1.57 |
| 8 | Q5 | 3.20 | 2.84 | 4.36 | 3.21 | 0.89 | 1.36 | 1.54 |
| 9 | S3 | 1.20 | 1.99 | 1.82 | 1.73 | 1.67 | 1.52 | 0.91 |
| 10 | Q6 | 6.20 | 4.81 | 8.85 | 5.93 | 0.78 | 1.43 | 1.84 |
| 11 | S4 | 1.66 | 1.45 | 3.09 | 1.80 | 0.88 | 1.87 | 2.13 |
| 12 | Q7 | 8.09 | 7.35 | 14.60 | 8.85 | 0.91 | 1.80 | 1.99 |
| 13 | Q8 | 4.86 | 4.67 | 6.24 | 5.00 | 0.96 | 1.28 | 1.34 |
| 14 | Q9 | 7.26 | 4.25 | 5.59 | 5.37 | 0.59 | 0.77 | 1.32 |
| 15 | Q10 | 7.73 | 10.38 | 7.29 | 9.05 | 1.34 | 0.94 | 0.70 |
| 16 | Q11 | 4.03 | 4.57 | 21.71 | 7.44 | 1.13 | 5.39 | 4.75 |
| 17 | Q12 | 3.55 | 6.63 | 13.50 | 6.93 | 1.87 | 3.80 | 2.04 |
| 18 | S5 | 2.14 | 1.71 | 2.69 | 2.01 | 0.80 | 1.26 | 1.57 |
| 19 | Q13 | 6.10 | 5.53 | 7.04 | 5.97 | 0.91 | 1.15 | 1.27 |
| 20 | Q14 | 3.49 | 3.23 | 7.67 | 4.09 | 0.92 | 2.19 | 2.37 |
| 21 | Q15 | 3.02 | 3.16 | 3.96 | 3.26 | 1.05 | 1.31 | 1.25 |
| 22 | Log out | 6.48 | 5.86 | 6.90 | 6.22 | 0.90 | 1.07 | 1.18 |
| | Mean | 149.81 | 145.43 | 227.07 | 161.12 | 0.97 | 1.52 | 1.56 |

The results of group B compared with the results of group A based on the divided average time show that for 38.1% of cases (8/21) time for B was longer with a value greater than 1; however, the mean value of this relation, 0.97, shows that the results for groups A and B were similar as was earlier confirmed by Mann-Whitney $U$ test (see Table 3). Differences between groups C and A and C and B are more clear with longer mean times of 1.52 and 1.56, respectively. A comparison of C versus A shows that, in 19/22 cases (90.48%), time was longer for participants in group C. Comparison of C versus B shows that, in 20/22 cases (90.91%), time was longer for participants in group C.

Performance for sections S1, S2, S3, S4, and S5 as expected was related to the number of questions and the individual complexity. An average time to answer all questions within section S1 for all users was 36.06 seconds. The time for groups A and B was similar with values of 31.05 and 32.50 seconds, respectively, while the time for group C was the highest with a value of 55.07 seconds. The average time to answer questions within S1 was 12.02 seconds per question, with 10.35 seconds for group A, 10.83 seconds for group B, and 18.36 seconds for group C. Time per question for the participants in group C was 1.77 times higher than the results from group A and 1.69 times higher than the users from group B. The first section of the questionnaire was the first contact with the application and users were familiarizing with the interface of the application, the way the virtual keyboard opens and closes, and they were accustomed to answering questions through the selection of answers from the list. The long average time was the result of a new, not used before interface and the structure of the application. The users became accustomed to the interface, and after the first section, the next sections were finished with better performance.

Answering two questions within section S2 took on average 8.05 seconds. An average value of 8.37 seconds was registered for group A, while the lowest value of 6.95 seconds was observed for the users in group B. The highest value of 10.83 seconds was observed for group C. Results for group C were 1.56 times longer than the results of group B and 1.29 longer than the average time of group A. Answering each question took an average of 4.03 seconds, and the time was shortest for the participants from group B (3.48 sec.) and longest for the users from group C with a value of 5.42 seconds. The participants from group B had an average of 4.19 seconds per question. Section S3 with one question was completed with an average time of 5.93 seconds. The shortest time was observed for the users from group B (4.81 seconds), while the users from group C required an average time of 8.85 seconds. This was 1.84 times longer than the participants from group B and 1.43 times longer than the participants from group A. The average time for the participants in group A was 6.20 seconds. Section S4 had the largest number of questions (6), and groups A and B achieved very similar results with the average times needed for answering all the questions being equal to 35.52 and 37.85 seconds, respectively. The average time for the participants from group C was equal to 68.93 seconds. The average time for all groups to answer a single question was 7.11 seconds. The highest value was equal to 11.49 seconds (group C) and that was 1.94 times longer than in group A (5.92 sec.) and 1.82 times longer than in group B (6.31 sec.). Similarly, the longest time to finish all three questions in section S5 was observed for the users from group C with 18.67 seconds required, while the time was equal to 12.61 sec. and 11.92 seconds for groups A and B, respectively. The average time per question for group C was 6.22 seconds and that was 1.48 times higher than for group A and 1.57 times higher than for group B.

Results did not show differences between groups A and B. For group A, the time was slightly shorter for sections S1 and S4 than for group B, while the reverse relationship was observed for other sections of the questionnaire. A stable pattern was observed for the users from group C with lower performance than the users from groups A and B in all cases. The differences between the participants from group C and from group A were higher for sections with higher numbers of questions. For sections S1 (3 questions), S4 (6 questions), and S5 (3 questions), the differences were 1.77, 1.94, and 1.48 times longer, respectively, while for sections S3 (one question) and S2 (two questions) the differences were longer at 1.43 and 1.29 times. Similar relationships were observed between the users from group C and group B. The average performance per question answered was longer for sections with high numbers of questions for all sections apart from S1 because of the longer time for processing because of initial learning. The average time per question was 7.11 seconds for section S4. Average time was shorter for sections S2, S3, and S5 and the values 4.03, 5.93, and 4.44 were obtained for questions 2, 1, and 3, respectively. However, a scalable relationship was not observed for time growing proportionally to the number of questions. Results are related to earlier findings that performance of tasks execution declines for the older adults [48]. Decreased information processing speed is observed within assumptions of Processing Speed Theory and the fact that cognitive acts for the older adults take longer and are more difficult to perform at all stages of the task [49].

### 5.4. Eye Tracking Based Analysis of Behaviours

*5.4.1. Patterns in Group A.* The main goal of eye tracking based study was to identify differences in the behaviour of the users in the three groups affecting the performance and the potential drawbacks or factors negatively affecting the user experience. Detailed results are based on the individual behaviours. Heatmaps A1–A6 in Figure 6 illustrate the behaviours observed in group A for the users of the ages ranging from 18 to 29. This group represents the users who are familiar with the mobile technologies, and the usage of the interface did not result in major problems with an average time of 149.8 seconds for answering all the questions. Most of the tasks within the questionnaire were done with the high performance. Eye tracking revealed elements of the interface interfering with the scanning paths and showed minor problems that needed to be fixed in the next stage of the development of the application. For example, the position of the lists with selectable answers in some cases was covering the question, and it was not clear how it was affecting the usage. Recorded eye movements and heat maps representing them illustrate the confusion caused by this problem (A1). The users were focusing attention on the question above the list while it was related to the earlier field, not to the currently selected list. Reading the unrelated question took 2.62 seconds, and after that attention was focused on the correct list. The whole time spent on the list related to the education level was 6.37 seconds. Another pattern shows how moving to another screen could be improved (A2). Localization of the NEXT button in the lower part of the screen resulted in a longer scanning path through the empty space in section S1. After switching to section S2, the attention was still concentrated on the empty space in the lower part of the screen, while questions were located at the top part of the display. The position of the NEXT button moved to the top part of the screen would help to avoid longer scanning path on both parts of interface.

Apart from the minor drawbacks, some typical behaviours were observed related to the quick scanning of the content and the absorption of the content with the peripheral vision (A3–A6). The users in group A did not experience major troubles with the usage of the virtual keyboard appearing from the first field in the form related to the number of persons in the household. For example, it took 4.84 seconds to get the number entered after loading section S1 of the questionnaire (A3). Finding the keys with the numerical values was done without high focus on the keyboard using peripheral vision, and the user switched off the keyboard properly. Even though section S4 with six questions contrasted with earlier parts of application with not fully used space, it showed that the user was not confused after switching to this section and was task oriented with attention focused on a single question (A4). For the textual content, the peripheral vision without focusing on the whole sentences was the main pattern (A5). Quick scanning of content took place with fast eye movement to choose the correct answer for the question, and it was supported by the peripheral vision (A6).

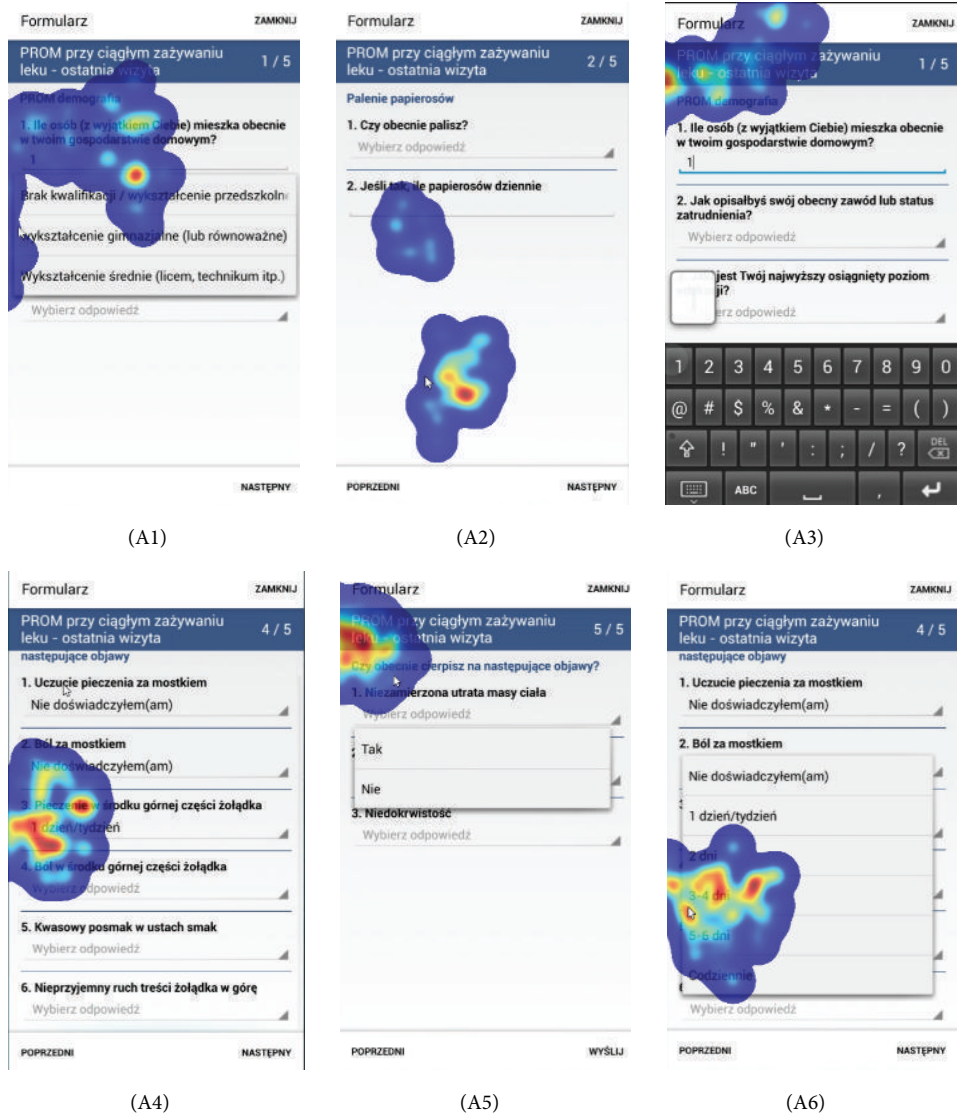

Figure 6: *Patterns from group A.* (A1) Unnecessary and misleading reading of the not related question above the list; (A2) attention kept on the empty space after switching to another section of the questionnaire; (A3) the correct usage of the keyboard and fast moving focus to the keyboard without fixations between starting and ending points; (A4) the attention concentrated on a single question and task on a multiquestion form; (A5) the adoption of the fact that the questions above the list are not necessarily related to the currently opened list; (A6) the peripheral vision used for reading the questions when the list with the answers is open.

*5.4.2. Patterns in Group B.* For the users in the second group of the ages ranging from 29 to 50, the average time to fill the questionnaire was 145.43 seconds. Some drawbacks related to the usage of the keyboard were observed as is illustrated in Figure 7 (B1–B3). The user was confused with the use of the virtual keyboard on the first questionnaire field. It took 11.02 seconds from loading the form to select the number of persons in the household (B1-B2). When all three questions were answered, the keyboard was not closed, and it was difficult to locate the NEXT button. The user was scanning the whole screen to localize it, while it was hidden by the keyboard (B3). Finally, the user took the necessary action and closed the keyboard (B4) after 14.17 seconds. Other patterns confirm the experience of the user with no further problems identified. After the keyboard was closed, the user analysed the new situation on the screen very quickly and focused on the area with the NEXT button (B4). When the user was moving the attention away from the keyboard to focus on another question, parts of the section were not observed (B2).

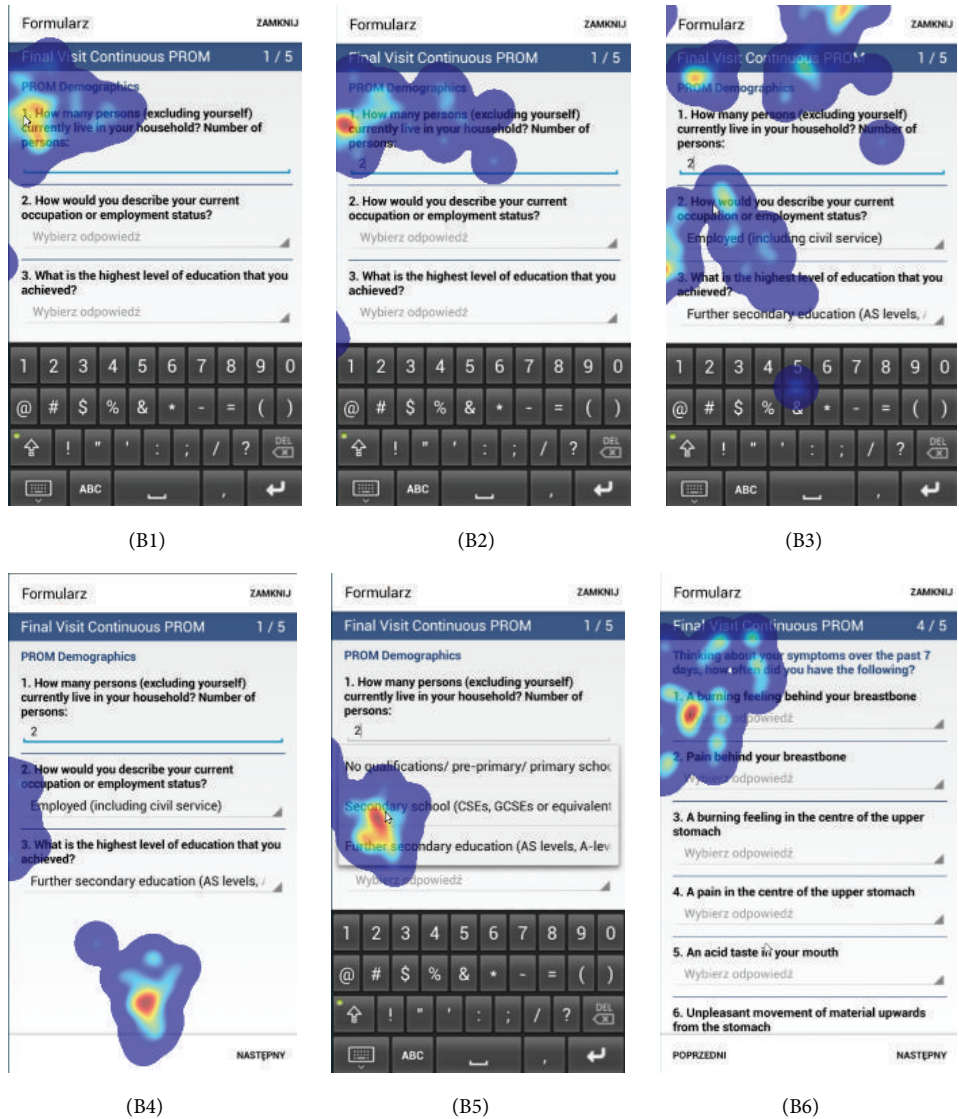

Figure 7: *Patterns from group B.* (B1) The use of the peripheral vision without detailed scanning of the whole screen; (B2) the task oriented user without unnecessary fixations when moving eyes to a keyboard; (B3) the confusion observed while the NEXT button is not visible; the user is looking without success at characteristic parts of the screen to find the desired option; (B4) after the keyboard disappears the user focuses the attention on the hidden keyboard screen and very quickly is analysing a new situation with quick focus on the area with the NEXT button; (B5) the attention is precisely focused only on the list and not at the parts above the list; (B6) Section 4/5 with a high number of the questions is not affecting the attention, and it is focused on the first question only.

The attention was focused only on the list (B5) not on the parts above. The usage of the peripheral vision with fixations observed only at the first part of the questions was visible in most cases (B1, B2, B5, and B6). The peripheral vision was used for both single and two-line texts, with the focus observed at the first part of the text. Section S4 with a high number of questions did not affect the attention, and the attention was focused on the first question as a current task to do (B6).

*5.4.3. Patterns in Group C.* While the behaviours and time to fill the questionnaire were similar for the users in groups A and B, the average time for the participants from group C was higher by 56% and equal to 227.07 seconds. The usage patterns and behaviours for the users of ages ranging from 51 to 69 are illustrated in Figure 8 (C1–C12).

Within section S1, the keyboard was automatically showed and the users had difficulty with the correct selection of the numerical values from the keyboard (C1). It took 18.97 seconds from loading the section until the numerical values appeared in the first field. After answering the third question, the users were locked and did not know how to switch to the next section of the questionnaire. While for the users in group B, in a similar situation, a quick scanning of the screen was observed, the users from group C concentrated attention on the middle part of the screen without an intuitive

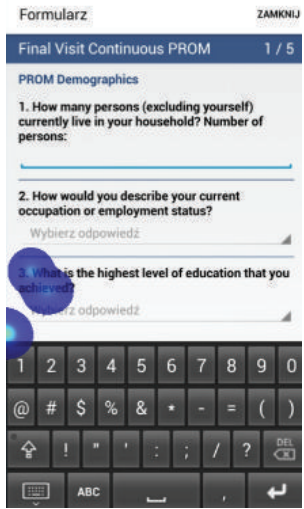
(C1)

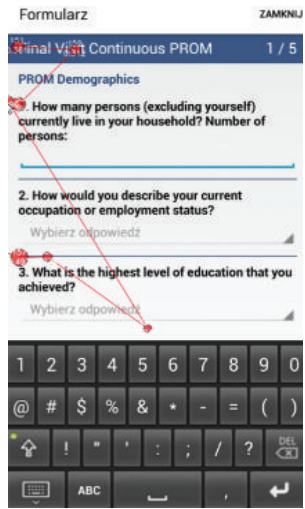
(C2)

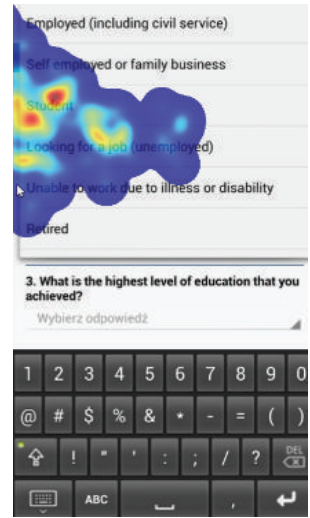
(C3)

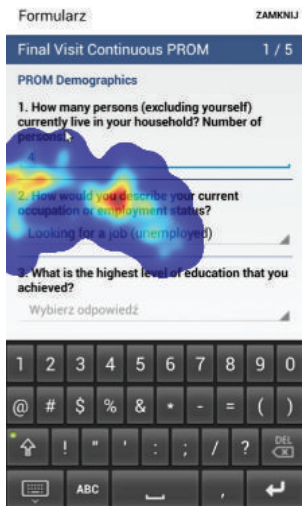
(C4)

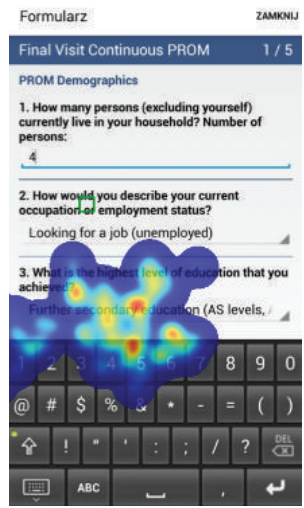
(C5)

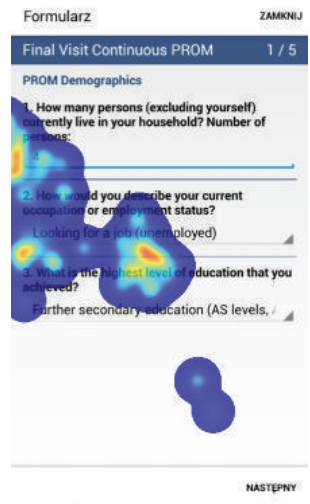
(C6)

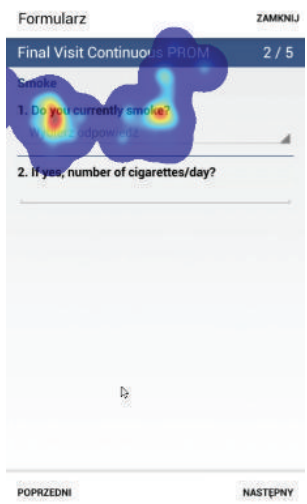
(C7)

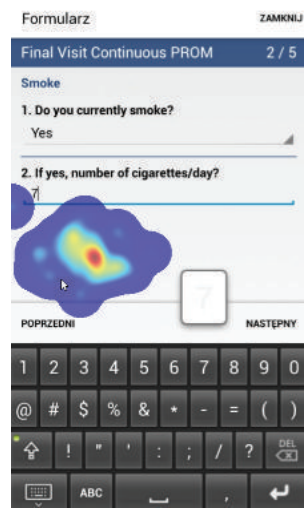
(C8)

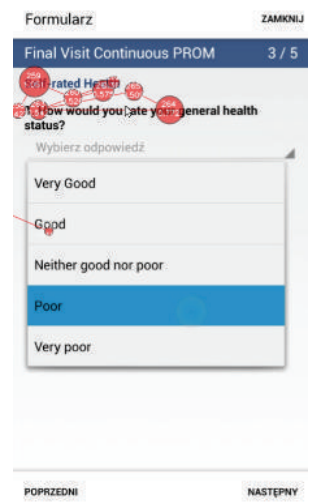
(C9)

Figure 8: Continued.

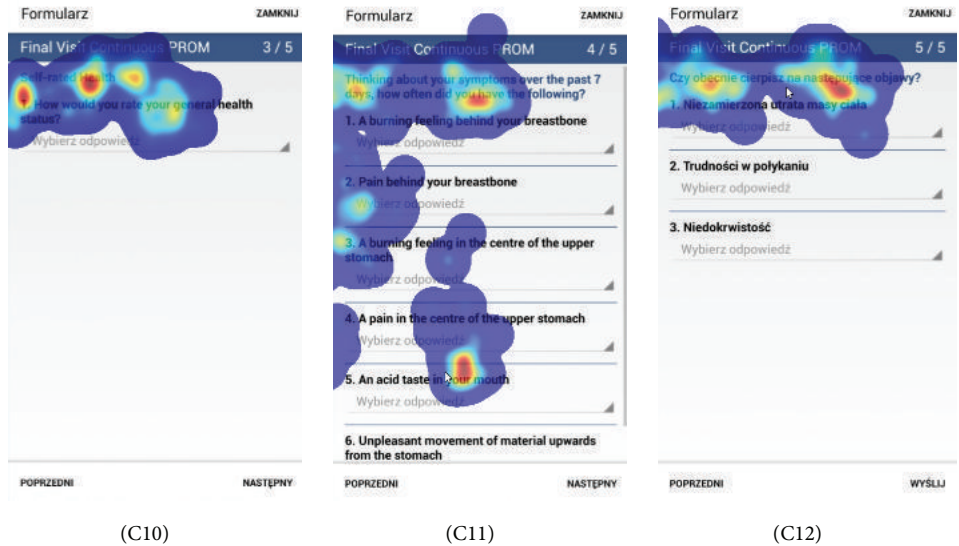

(C10)  (C11)  (C12)

Figure 8: *Patterns from group C.* (C1) The virtual keyboard created confusion and was opened and closed twice before entering the correct number was introduced; (C2) the attention focused in the middle of the starting and target point; (C3) the higher attention is put on the questions on the list with complete reading; (C4) reading the questions with the higher attention; (C5) a long-lasting confusion observed when the keyboard covers the NEXT button; (C6) after closing the keyboard, a new situation on the screen is analysed and the attention is put back on the questions in the search for the NEXT button after a long time; (C7) a careful reading of the questions before launching the list with the answers; (C8) a more intuitive and quick keyboard usage on section S2; (C9) double scanning the questions within return path; (C10) the peripheral vision is less used and the questions are followed carefully; (C11) after switching to section S4 with a higher number of the questions, the attention is spread among several parts of the screen before the primary task is continued; (C12) a careful reading of the questions without the use of the peripheral vision continues.

search for the desired option (C5). Finally, as soon as the keyboard disappeared (C6), the users spent 16.30 seconds analysing the situation on the screen and searching for the way to go to the next section (C6). Differences in behaviours for the users in group C were observed when compared to groups A and B. The eye paths lasted longer and the users focused vision in the middle of the paths before the initial and starting target points. For example, when moving attention to the other sections, the keyboard fixation is observed in the middle (C2). The users were reading the whole answers more slowly with the lower usage of peripheral vision than the users from groups A and B. The higher attention was placed on reading in detail the questions on the list (C3, C4, C6, C10, and C12). However, after the initial failure with the usage of the keyboard, the user had no problems with the usage of the keyboard in section S2, and it took 4.48 seconds to select the values on the numerical keyboard from the time the question was displayed (C8). Careful reading of the questions resulted in returning paths and full text scanning with the central vision even twice (C9). A different behaviour was observed for section S4 with the higher number of the questions than for the users in groups A and B earlier (C11). The users were looking at all the questions slowly before starting to answer question number one. Overall, the differences in the behaviours between group C and the other groups showing slower reading and the worse control of eye movements of older participants. This is consistent with earlier studies [50, 51].

*5.5. Google Analytics.* The Google Analytics framework is a useful tool for monitoring the user behaviour within the application or on the webpage. In the TRANSFoRm project, the tool was implemented in the Android application and gathers the following information:

(i) The amount of the time spent on the particular screen.
(ii) The number of the successful and the unsuccessful log in and log out attempts.
(iii) The number of the opened and the abandoned questionnaires.
(iv) The number of incomplete answers (when the users have missed a question).
(v) The number of times when the session has expired.
(vi) The number of the button clicks of all buttons present in the application.

The idea behind gathering such statistics is to detect the potential problems when using the application and to fix them before the users reports the problem.

For example, the large number of clicks on the "Forgot Password" button may suggest that the generated passwords are too complicated for the users. The large number of missing answers may mean that there are too many questions on the screen or some of them are not properly separated from the others and the users do not see them. The great number of expired sessions may suggest that the users need more time for the interaction with the application and the session time should be extended.

It is also possible to improve the layout based on the Google Analytics statistics. For example, very small number of clicks on the "Previous" button might be a suggestion to remove this button completely.

Currently, the number of gathered statistics is too small to present it and to provide more concrete conclusions.

## 6. Conclusions

The TRANSFoRm Study System was designed as a generic solution to support embedding of clinical trial functionality into Electronic Health Record systems and providing electronic data collection capabilities. The only requirement is that the study is designed in accordance with the CDISC (Clinical Data Interchange Standards Consortium) SDM (Study Design Model) standard [52]. If this condition is fulfilled and the SDM standard is used then all of the ODM questionnaires can be used to display questionnaires on the mobile and web devices to the patient.

The entire TSS was GCP (Good Clinical Practice) certified and validated with the participation of 10 patients [5]; however, neither of those procedures revealed any substantial information on how to improve the mobile applications. That is why additional evaluation of mobile applications was performed.

Quantitative and qualitative research delivered useful feedback for further mobile applications development. The application uses standard techniques used in mobile applications based on Android and iOS devices. The more experienced users familiar with mobile technologies from groups A and B had no major problems. The users were performing the tasks very quickly with the use of very fast scanning of the content visible on the screen. They were task oriented with focusing on the task to do instead of scanning the whole area of the screen and the application. However, some minor drawbacks were observed in regard to the position of the list with the answers to be selected. The virtual keyboard created some problems while it opened unexpectedly and automatically. For this type of application in which only limited values are possible, it would be possible to standardize the interface and use only selectors. The keyboard could be replaced with the list of values to select while there was no open question. When switching the screen, the positions of elements in the interface should be localized closer to the content and adjusted to the location of the content on the next screen to avoid searching for that option to go ahead to next section and searching for the main content in the newly opened section. The experienced users filter unnecessary information, and they avoid reading unnecessary content during different stages of the task. This was observed when new parts of applications were shown.

The participants from all groups showed high learning abilities and after filling out the first section with low performance the next sections were filled without the problems. The lower performance was observed among participants with the age above 50, but the users from this group, after filling the first section, had no problems with the other parts of the questionnaire. The training session and presentation of mechanics could be done to improve older users' experiences and limit the frustrations related to the new type of the interface. Also other options can be implemented, such as tutorial or an interactive user guide.

## Competing Interests

The authors declare that they have no competing interests.

## Acknowledgments

This work was partially supported from the European Union's Seventh Framework Programme for research, technological development, and demonstration under Grant Agreements no. 247787 (TRANSFoRm) and no. 316097 (ENGINE) and by the Faculty of Computer Science and Management, Wrocław University of Science and Technology statutory funds.

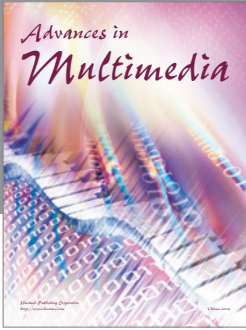
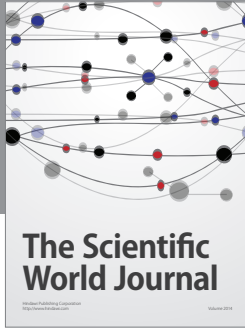
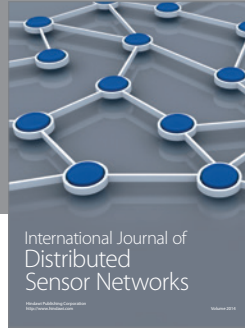
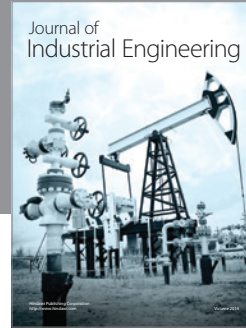
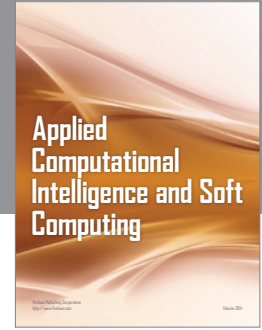
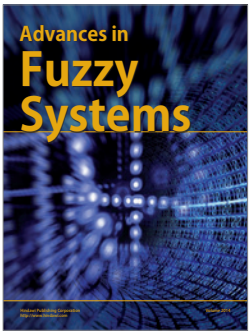
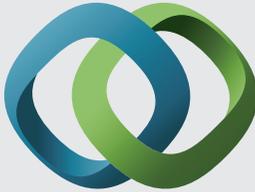
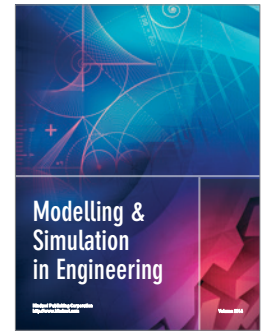
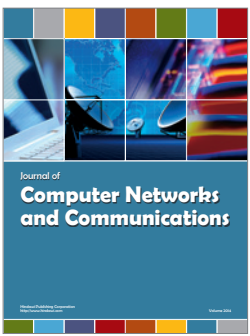
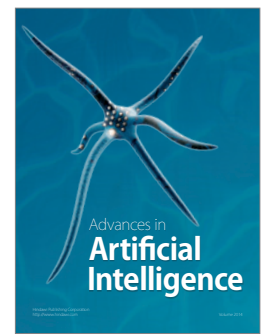
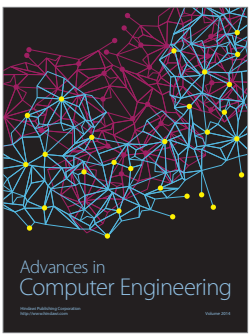
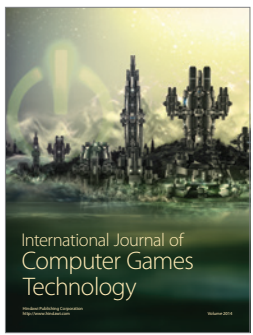
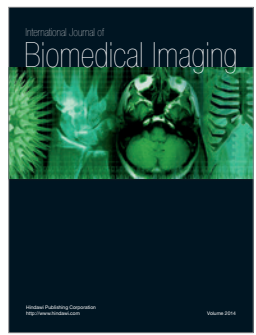
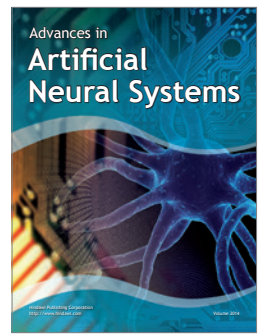
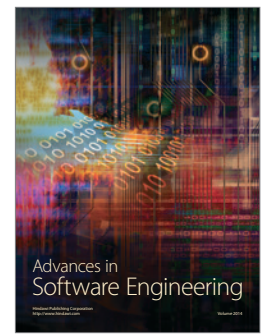
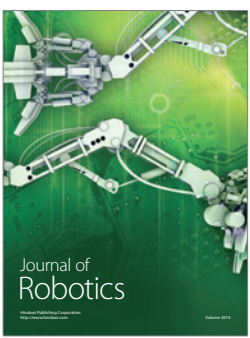
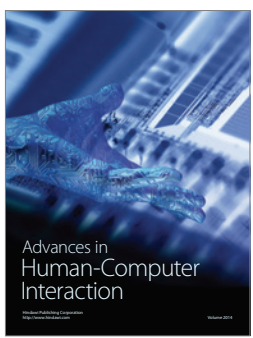
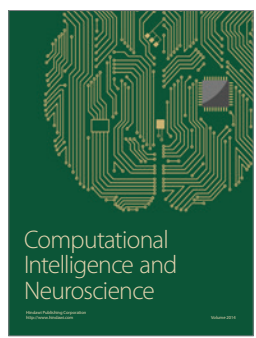
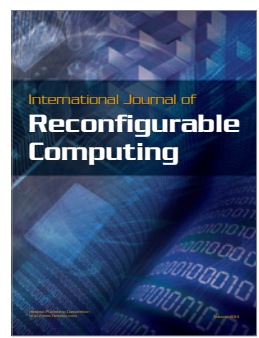
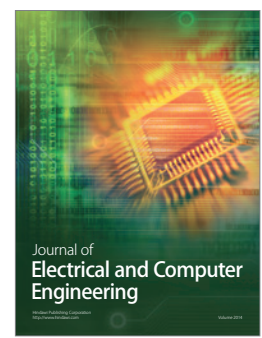